\documentclass[letterpaper]{article}
\usepackage[ascii]{inputenc}
\usepackage[T1]{fontenc}
\usepackage[english]{babel}
\usepackage{amsmath}
\usepackage{amssymb,amsfonts,textcomp}
\usepackage{color}
\usepackage[top=1in,bottom=1in,left=0.75in,right=0.75in,nohead,nofoot]{geometry}
\usepackage{multicol}
\usepackage{array}
\usepackage{hhline}
\usepackage{hyperref}
\hypersetup{colorlinks=true, linkcolor=blue, citecolor=blue, filecolor=blue, urlcolor=blue}
\makeatletter
\newcommand\arraybslash{\let\\\@arraycr}
\makeatother
\makeatletter
\newcommand\ps@Standard{
  \renewcommand\@oddhead{}
  \renewcommand\@evenhead{}
  \renewcommand\@oddfoot{}
  \renewcommand\@evenfoot{}
  \renewcommand\thepage{\arabic{page}}
}
\makeatother
\title{Partition Parameters for Girth Maximum  $(m,r)$ BTUs}
\begin{document}
\clearpage\setcounter{page}{1}\pagestyle{Standard}
{\centering\bfseries
\Large{\textbf{Partition Parameters for Girth Maximum  $(m,r)$ BTUs}}
\par}

{\centering\itshape
Vivek S Nittoor\ \ \ \ Reiji Suda
\par}

{\centering
Department Of Computer Science, The University Of Tokyo
\par}

{\centering
Japan
\par}

\bigskip

\bigskip

\bigskip

\bigskip
\begin{multicols}{2}
{\bfseries
\textit{Abstract}{}---This paper describes the calculation of the optimal partition parameters such that the \ girth maximum $(m,r)$  Balanced Tanner Unit \ lies in family of BTUs specified by them \ using a series of proved results and thus creates a framework for specifying a search problem for finding the girth maximum $(m,r)$  BTU. Several open questions for \ girth maximum $(m,r)$  BTU have been raised.}

{\bfseries\itshape
Keywords Permutation Groups, Cycle Index, Balanced Tanner Unit, Girth Maximum $(m,r)$ BTU}

\section{Introduction}
We have introduced a family of bi-partite graphs called Balanced Tanner Units (BTUs) \ in $[1]$ and have introduced a family of graphs  $\Phi (\beta _{1},\beta _{2},\ldots ,\beta _{r\text{--}1})$ . The goal of this paper is to derive the forms for optimal partitions  $\beta _{1},\beta _{2},\ldots ,\beta _{r\text{--}1}$ such that the girth maximum  $(m,r)$ BTU lies in  $\Phi (\beta _{1},\beta _{2},\ldots ,\beta _{r\text{--}1})$ using a series of mathematical results that build upon the fundamentals of BTUs introduced in $[1]$ . 

We review a few definitions from  $[1]$.

\subsection[Set]{Set $P_{2}(m)$ }
 $P_{2}(m)$ refers to the set of partitions of  $m\in \mathbb{N}$ that consist of numbers that are greater than or equal to $2$.

\subsection{Partition Component }
If  $\beta \in P_{2}(m)$ refers to  $\sum _{j=1}^{y}q_{j}=m$ then \ each  $\{q_{j}\}$ for  $1\le j\le y$ is referred to as a partition component of  $\beta $. 

\subsection[Definition of]{Definition of $\Phi (\beta _{1},\beta _{2},\ldots ,\beta _{r\text{--}1})$ }
 $\Phi (\beta _{1},\beta _{2},\ldots ,\beta _{r\text{--}1})$ refers to the family of all labeled $(m,r)$ BTUs with compatible permutations  $p_{1,}p_{2,}\ldots ,p_{r}\in S_{m};p_{i}\notin C(p_{1},p_{2},\ldots ,p_{i-1})$ for  $1<i\le r$ that occur in the same order on a complete $m$ symmetric permutation tree ,  $x_{1,1}<x_{2,1}<\ldots <x_{r,1}$ where  $p_{j}=(x_{j,1}x_{j,2}\ldots x_{j,m});1\le j\le r$ , such that  $\beta _{i-1}$  is the partition between permutations $p_{i\text{--}1}$  and  $p_{i}$  for all integer values of  $i$  given by  $1<i\le r$ .

\subsection{Symmetric Permutation Tree and its properties}
{\selectlanguage{english}
A  $m$  Symmetric permutation tree  $S_{\mathit{PT}}\{m\}$  is defined as a labeled tree with the following properties:}

\begin{enumerate}
\item {\selectlanguage{english}
 $S_{\mathit{PT}}\{m\}$ has a single root node labeled  $0$ .}
\item {\selectlanguage{english}
 $S_{\mathit{PT}}\{m\}$ has $m$  nodes at \ depth  $1$ from the root
node.}
\item {\selectlanguage{english}
 $S_{\mathit{PT}}\{m\}$ has nodes at depths ranging from  $1$ to $m$ ,
with each node having a labels chosen from  $\{1,2,\ldots ,m\}$ . The
root node  $0$ has  $m$  successor nodes. Each node at \ depth  $1$ has
 $m-1$  successor nodes at depth  $2$ . Each \ node at depth  $i$  has 
$m\text{--}i+1$  successor nodes at depth $i+1$  . Each node at depth 
$m\text{--}1$  has  $1$  successor node at depth $m$  .}

\item {\selectlanguage{english}
No successor node in  $S_{\mathit{PT}}\{m\}$ has the same node label as
any of its ancestor nodes.}
\item {\selectlanguage{english}
No two successor nodes that share a common parent node have the same
label. }
\item {\selectlanguage{english}
The sequence of nodes in the path traversal from the node at depth  $1$ 
to the leaf node at depth  $m$  in  $S_{\mathit{PT}}\{m\}$  represents
the permutation represented by the leaf node.}
\item {\selectlanguage{english}
 $S_{\mathit{PT}}\{m\}$ has  $m!$  leaf nodes each of which represent an
element of the symmetric group of degree  $m$ denoted by  $S_{m}$  .}
\end{enumerate}

\subsection{Compatible Permutations }
\begin{enumerate}
\item {\selectlanguage{english}
Two permutations on a set of  $s$  elements represented by 
$(x_{1}x_{2}\ldots x_{s});x_{p}\neq x_{q}$
$\forall p\neq q;1\le p\le
s;1\le q\le s;p,q\in \mathbb{N}$  where  $1\le x_{i}\le s;i\in
\mathbb{N};1\le i\le s$ and  $(y_{1}y_{2}\ldots y_{s});y_{p}\neq y_{q}$
 $\forall p\neq q;1\le p\le s;1\le q\le s;p,q\in \mathbb{N}$  where 
$1\le y_{i}\le m;i\in \mathbb{N};1\le i\le s$ are compatible if and
only if  $x_{i}\neq y_{i}\forall i\in \mathbb{N};1\le i\le s$ .}
\item {\selectlanguage{english}
A set of  $r$  permutations on a set of  $s$  elements represented by 
$(x_{i,1}x_{i,2}\ldots x_{i,s});x_{i,p}\neq x_{i,q}$  $\forall p\neq
q;1\le p\le s;1\le q\le s;p,q\in \mathbb{N}$ where  $1\le x_{i,\alpha
}\le s\forall 1\le i\le r;1\le \alpha \le s;i,\alpha \in \mathbb{N}$
are compatible if and only if  $x_{i,\alpha }\neq x_{j,\alpha }$ 
$\forall i\neq j;1\le \alpha \le s;1\le i\le r;1\le j\le r;$  
$i,j,\alpha \in \mathbb{N}$ .}
\end{enumerate}
\subsection{Notation}
{\selectlanguage{english}
 $p_{i}\notin C(I_{m},p_{2},\ldots ,p_{i\text{--}1})$ :  $p_{i}$  is
compatible with permutations $I_{m},p_{2},\ldots ,p_{i\text{--}1}$ .}

\section[General Approach at Constructing BTU with the maximum girth]{General Approach at Constructing $(m,r)$  BTU with the maximum girth}
\subsection{Conjecture}
Any non-isomorphic $(m,r)$ BTU can be constructed by choosing compatible permutations  $p_{1},p_{2},\ldots ,p_{r}\in S_{m};p_{j+1}\notin C(p_{1},p_{2},\ldots ,p_{j})$ for  $1\le j\le r-1$ that occur in the same order on a  $m$ complete symmetric permutation tree,  $x_{1,1}<x_{2,1}<\ldots <x_{r,1}$ i.e.,  $x_{j,1}<x_{j+1}$ for  $1\le j\le r-1$ , where $p_{i}=(x_{i,1}x_{i,2}\ldots x_{i,m})$ for  $1\le i\le r$.

\textbf{Proof} Without loss of generality, any labeled $(m,r)$ BTU can be represented by a choice of permutations $p_{1},p_{2},\ldots ,p_{r}\in S_{m}; p_{j+1}\notin C(p_{1},p_{2},\ldots ,p_{j})$ for  $1\le j\le r-1$ that occur in the same order on a  $m$ complete symmetric permutation tree,  $x_{1,1}<x_{2,1}<\ldots <x_{r,1}$ and hence any non-isomorphic  $(m,r)$ BTU can be constructed by choosing compatible permutations  $p_{1},p_{2},\ldots ,p_{r}\in S_{m};p_{j+1}\notin C(p_{1},p_{2},\ldots ,p_{j})$ for  $1\le j\le r-1$ that occur in the same order on a  $m$ complete symmetric permutation tree, $x_{1,1}<x_{2,1}<\ldots <x_{r,1}$.

\subsection{Conjecture}
Any non-isomorphic $(m,r)$ BTU can be constructed by choosing compatible permutations  $p_{1},p_{2},\ldots ,p_{r}\in S_{m};p_{j+1}\notin C(p_{1},p_{2},\ldots ,p_{j})$ for  $1\le j\le r-1$ such \ that  $x_{j,1}=j\mathit{for}1\le j\le r$, where $p_{i}=(x_{i,1}x_{i,2}\ldots x_{i,m})$ for  $1\le i\le r$.

\textbf{Proof} Without loss of generality, any non-isomorphic  $(m,r)$ BTU is isomorphic to a BTU represented by a set of permutations $\{p_{1} ,p_{2}, \ldots ,p_{r}\}$ where \  $p_{1},p_{2},\ldots ,p_{r}\in S_{m};p_{j+1}\notin C(p_{1},p_{2},\ldots ,p_{j})$ for  $1\le j\le r-1$ such \ that  $x_{j,1}=j$ for $1\le j\le r$, since it could be brought to the form by row exchanges or permutations on depth for the permutation representation of the labeled  $(m,r)$ BTU.

\subsection[Theorem ]{Theorem }
Any  $(m,r)$  BTU is isomorphic to an element of  $\Phi (\beta _{1},\beta _{2},\ldots ,\beta _{r\text{--}1})$ for some choice of \  $\beta _{1},\beta _{2},\ldots ,\beta _{r\text{--}1}\in P_{2}(m)$ .

\textbf{Proof} \ In general any  $(m,r)$  BTU can be diagonalized by suitable row and column exchanges, and without loss of generality,  $p_{1}=I_{m}$  \ .We map the positions of $1$ s in the first column to  $x_{i,1};2\le i\le r$  where  $p_{i}=(x_{i,1}x_{i,2}\ldots x_{i,m})$ . Thus, map the \  $(m,r)$  BTU to compatible permutations  $p_{1},p_{2},\ldots ,p_{r}$  that occur in the same order on a complete  $m$  symmetric permutation tree. Let  $\beta _{i}\in P_{2}(m)$  be the partition represented between  $p_{i}$  and  $p_{i+1}$  for  $1\le i\le r\text{--}1$ . Hence, any $(m,r)$  BTU is isomorphic to an element of  $\Phi (\beta _{1},\beta _{2},\ldots ,\beta _{r\text{--}1})$ for some choice of \  $\beta _{1},\beta _{2},\ldots ,\beta _{r\text{--}1}\in P_{2}(m)$ . 

\subsection{Implicit Enumeration Conjecture }
All non-isomorphic  $(m,r)$  BTUs can be enumerated by 

\begin{enumerate}
\item An exhaustive enumeration of $\{\beta _{1},\beta _{2},\ldots ,\beta _{r\text{--}1}\}$ where $\beta _{1},\beta _{2},\ldots ,\beta _{r\text{--}1}\in P_{2}(m)$.
\item For each of the above enumeration of  $\{\beta _{1},\beta _{2},\ldots ,\beta _{r\text{--}1}\}$ we construct all non-isomorphic  $(m,r)$ BTU with  $\beta _{i}$  being the partition between  $p_{i\text{--}1}$  and  $p_{i}$  for all integer values of  $i$  given by  $2\le i\le r-1$ represented by the set $\Phi (\beta _{1},\beta _{2},\ldots ,\beta _{r\text{--}1})$ .
\end{enumerate}
\subsection{Definition of Micro-partition }
{\bfseries
\textmd{Given partitions } $\beta _{1},\beta _{2},\ldots ,\beta _{r-1}\in P_{2}(m)$ \textmd{with } $\beta _{i}$ \textmd{of the form } $\sum _{j=1}^{y_{i}}p_{i,j}=m;i\in \mathbb{N},1\le i\le r-1$ \textmd{, a micro-partition of } $\beta _{i+1}$ \textmd{with respect to } $\beta _{i}$ \textmd{is defined as } $x_{i,j,z}\in \mathbb{N}\cup \{0\};1\le i<r-1;1\le j\le y_{i};1\le z\le y_{i+1}$ \textmd{ such that} $\sum _{z=1}^{y_{i+1}}x_{i,j,z}=p_{i,j};1\le j\le y_{i};1\le i<r-1$ \textmd{and} $\sum _{j=1}^{y_{i}}x_{i,j,z}=p_{i+1,z};1\le z\le y_{i+1};1\le i<r-1$ \textmd{.}}

\subsection{Micro-partition to label mapping}
{\bfseries
\textmd{For each micro-partition of } $\beta _{i+1}$ \textmd{with respect to } $\beta _{i}$ \textmd{, } $x_{i,j,z}\in \mathbb{N}\cup \{0\};1\le i<r-1;1\le j\le y_{i};1\le z\le y_{i+1}$ \textmd{we define }Micro-partition to label mapping $y(x_{i,j,z})$ \textmd{as an ordered set of } $x_{i,j,z}$ \textmd{distinct labels. }}

\begin{enumerate}
\item For  $\beta _{1}$ , the labels from  $\{1,2,\ldots ,m\}$  for each partition component gets fixed by $\Psi (\beta _{1})$.
\item For  $\beta _{2},\ldots ,\beta _{r-1}$ each distinct choice of a micro-partition to label mapping leads to numerous labeled graphs, which in turn would correspond to elements in  $\Phi (\beta _{1},\beta _{2},\ldots ,\beta _{r-1})$ .
\end{enumerate}
\subsection[Algorithm to enumerate all non{}-isomorphic BTUs for  ]{Algorithm to enumerate all non-isomorphic  $(m,r)$ BTUs for  $r\ge 3$ }
\begin{enumerate}
\item Enumerate all distinct ordered combinations of  $\{\beta _{1},\beta _{2},\ldots ,\beta _{r\text{--}1}\}$  where  $\beta _{1},\beta _{2},\ldots ,\beta _{r\text{--}1}\in P_{2}(m)$ .
\item For each instance of  $\{\beta _{1},\beta _{2},\ldots ,\beta _{r\text{--}1}\}$ , we enumerate all unique combinations of Micro-partitions of  $\beta _{i+1}$  w.r.t.  $\beta _{i}$ for  $1\le i\le r\text{--}1$ . 
\item For each instance of a combinations of Micro-partitions of  $\beta _{i+1}$  w.r.t.  $\beta _{i}$ for  $1\le i\le r\text{--}1$ , we enumerate all sets of  $r\text{--}1$  Unordered labeled partitions.
\item For each instance of a set of  $r\text{--}1$  Unordered labeled partitions, we enumerate all possible sets of ordered $r\text{--}1$ Ordered labeled partitions using Reduced Cycle Enumeration corresponding to each Unordered labeled partition enumerated in the previous step.
\item We construct $p_{2},p_{3},\ldots ,p_{r}$  corresponding to the set of $r\text{--}1$ Ordered labeled partitions enumerated in the previous step. ( $p_{1}=I_{m}$ ).
\end{enumerate}
We shall develop the rationale for the this algorithm in subsequent sections. 

\subsection[Algorithm ]{Algorithm  $\alpha $ }
A \ labeled  $(m,r)$  BTU can be characterized by a set of compatible permutations  $p_{1},p_{2},\ldots ,p_{r}\in S_{m}$ such that  $p_{i}\notin C(p_{1},\ldots ,p_{i-1});1<i\le r$ .

 $p_{1}=I_{m}$ ; \\
for(each possible  $p_{2};p_{2}\notin C(p_{1})$ ) \{ \\
for(each possible  $p_{3};p_{3}\notin C(p_{1},p_{2})$ ) \{

\ldots 
\\for(each possible  $p_{r};p_{r}\notin C(p_{1},p_{2},\ldots ,p_{r-1})$ ) \{
\\for(each chosen  $p_{1},p_{2},\ldots ,p_{r}\in S_{m}$ such that  $p_{i}\notin C(p_{1},\ldots ,p_{i-1});1<i\le r$)  \\We evaluate the girth of the chosen  $(m,r)$  BTU;
\\ \}
\\ \ldots 
\\ \}
\\ \}

Choose  $p_{1},p_{2},\ldots ,p_{r}\in S_{m}$ with the best girth from the above explorations. 

\subsection{Theorem }
{Algorithm  $\alpha $ chooses a  $(m,r)$  BTU with the best girth.}
\textbf{Proof} Even though the algorithm does not enumerate all labeled  $(m,r)$  BTUs since we choose  $p_{1}=I_{m}$ , it clearly covers the space of all non-isomorphic  $(m,r)$ BTUs \ with a huge amount of duplications since any labeled  $(m,r)$ BTU is an element of  $\Psi (\beta _{1},\beta _{2},\ldots ,\beta _{r-1})$ for some  $\beta _{1},\beta _{2},\ldots ,\beta _{r-1}\in P_{2}(m)$ . The algorithm clearly chooses a $(m,r)$  BTU with the best girth from elements of all possible sets  $\Phi (\beta _{1},\beta _{2},\ldots ,\beta _{r-1})$ . 

\subsection[Algorithm ]{Algorithm  $\alpha _{1}$ }
Enumerate all distinct possible  $\{\beta _{1},\beta _{2},\ldots ,\beta _{r-1}\}$ such that  $\beta _{1},\beta _{2},\ldots ,\beta _{r-1}\in P_{2}(m)$ ; 

for(each enumeration of  $\beta _{1},\beta _{2},\ldots ,\beta _{r-1}\in P_{2}(m)$) \{ 

 $p_{1}=I_{m}$ ;

for(each possible  $p_{2};\beta _{1}(p_{1},p_{2});p_{2}\notin C(p_{1})$ ) \{

for(each possible  $p_{3};\beta _{2}(p_{2},p_{3});p_{3}\notin C(p_{1},p_{2})$ ) \{

\ldots
\\ for(each possible  $p_{r};\beta _{r-1}(p_{r},p_{r-1});p_{r}\notin C(p_{1},p_{2},\ldots ,p_{r-1})$ ) \{
\\ We evaluate the girth of the chosen  $(m,r)$  BTU represented by the chosen permutations $p_{1},p_{2},\ldots ,p_{r}\in S_{m}$ such that  $p_{i}\notin C(p_{1},\ldots ,p_{i-1});1<i\le r$;

\}

\ldots \\
\}  \\
\} \\
\}

Choose  $p_{1},p_{2},\ldots ,p_{r}\in S_{m}$ with the best girth from the above explorations. 

\subsection{Theorem }
{\textup{Algorithm } $\alpha _{1}$ \textup{chooses a } $(m,r)$ \textup{ BTU with the best girth in } $\Phi (\beta _{1},\beta _{2},\ldots ,\beta _{r-1})$ \textup{given } $\beta _{1},\beta _{2},\ldots ,\beta _{r-1}\in P_{2}(m)$ \textup{.} }
\\ \textbf{Proof} \ The Algorithm  $\alpha _{1}$ clearly covers the space of all non-isomorphic  $(m,r)$ BTUs in the family \  $\Phi (\beta _{1},\beta _{2},\ldots ,\beta _{r-1})$ . Hence, the chosen  $p_{1},p_{2},\ldots ,p_{r}\in S_{m}$ by Algorithm $\alpha _{1}$ represents a  $(m,r)$  BTU with the best girth in the family \  $\Phi (\beta _{1},\beta _{2},\ldots ,\beta _{r-1})$ . 

\subsection{Ordered labeled Partition}
An ordered labeled Partition of  $m$  consists of distinct {ordered} $y$  subsets  $B_{1},B_{2},\ldots ,B_{y}$ of the set $A=\{1,2,\ldots ,m\}$  such that

\begin{enumerate}
\item  The ordered subsets satisfy the condition $B_{i}\cap B_{j}=\Phi $ the null set  $\forall i\neq j,1\le i\le y;1\le j\le y$ 
\item Each \textbf{ordered} set  $B_{i}\subset A$ with number of distinct elements  $p_{i};1\le i\le y$ . 
\end{enumerate}
\subsection{Unordered labeled Partition}
Given  $\beta \in P_{2}(m)$  which refers to  $\sum _{i=1}^{y}p_{i}=m$  , an {unordered }labeled partition of  $m$ is a collection of distinct $y$  subsets  $B_{1},B_{2},\ldots ,B_{y}$  of the set $A=\{1,2,\ldots ,m\}$  such that 

\begin{enumerate}
\item  $B_{i}\cap B_{j}=\Phi $ the null set $\forall i\neq j,1\le i\le y;1\le j\le y$ 
\item Each set  $B_{i}\subset A$ with number of distinct elements  $p_{i};1\le i\le y$ . 
\end{enumerate}
\subsection{Unordered labeled Partition Mapping Enumeration Problem}
Given  $\beta \in P_{2}(m)$  which refers to  $\sum _{i=1}^{y}p_{i}=m$  , the {unordered }labeled partition mapping Enumeration problem refers to the Enumeration of all distinct $y$  subsets  $B_{1},B_{2},\ldots ,B_{y}$  of the set $A=\{1,2,\ldots ,m\}$  such that 
\begin{enumerate}
\item  $B_{i}\cap B_{j}=\Phi $ the null set $\forall i\neq j,1\le i\le y;1\le j\le y$ .
\item Each set  $B_{i}\subset A$ with number of distinct elements  $p_{i};1\le i\le y$.
\end{enumerate}
\subsection[Ordered labeled Partition Mapping Enumeration Problem]{Ordered labeled Partition Mapping Enumeration Problem}
{\textup{Given } $\beta \in P_{2}(m)$\textup{ which refers to } $\sum _{i=1}^{y}p_{i}=m$\textup{ , the }\textbf{\textup{ordered }}\textup{labeled partition mapping Enumeration problem refers to the Enumeration of all distinct }\textbf{\textup{ordered}} $y$\textup{ subsets } $B_{1},B_{2},\ldots ,B_{y}$\textup{ of the }\textup{set} $A=\{1,2,\ldots ,m\}$\textup{ such that}}
\begin{enumerate}
\item  $B_{i}\cap B_{j}=\Phi $ ,the null set $\forall i\neq j,1\le i\le y;1\le j\le y$ .
\item Each ordered set $B_{i}\subset A$ with number of distinct elements  $p_{i};1\le i\le y$ . 
\end{enumerate}
\subsection[Partitions \ micro{}-partition Unordered labeled Partitions Ordered labeled Partitions  Compatible Permutations labeled BTUs]{Partitions \  $\to $ micro-partition  $\to $ Unordered labeled Partitions  $\to $ Ordered labeled Partitions  $\to $ Compatible Permutations  $\to $ labeled  $(m,r)$ BTUs}
Given a set of partitions $\beta _{1},\beta _{2},\ldots ,\beta _{r\text{--}1}\in P_{2}(m)$, we can have many possible sets of micro-partitions. For each set of micro-partitions, we can have many possible unordered labeled partitions. For each set of \ unordered labeled partitions, we can have many possible ordered labeled partitions. For each set of $r\text{--}1$ ordered labeled partitions, we can construct a set of compatible permutations

 $\{I_{m},p_{2},\ldots ,p_{r\text{--}1}\}$ which represents a labeled \  $(m,r)$ BTU.

\subsection{Multiple levels of enumeration}
{Depth Permutation \& Label permutations that preserve $p_{1}$  and  $p_{2}$ and create different instances of $p_{3}$ for the same}
\begin{enumerate}
\item Enumeration of all possible micro-partitions of  $\beta _{i+1}$  w.r.t. $\beta _{i}$ .
\item For each micro-partition, enumeration of all possible non-ordered labeled partitions corresponding to $\beta _{i+1}$ .
\item For each non-ordered labeled partition, enumerate all cycle orders.
\end{enumerate}
\subsection[Constrained labeled Partition Mapping Problem ]{Constrained labeled Partition Mapping Problem }
The Permutation Enumeration formulae derived in  $[1]$ gives us the number of candidate permutations corresponding to a specified partition. We recall from  $[1]$ that the number of distinct permutations $\{p_{2};\beta _{1}(p_{2},p_{1}) = (m) ,p_{2}\notin C(p_{1})\}$ is \ given by 




$f(\beta )=(m - 1) * \sum _{j,\mathit{distinct}p_{j}}{\frac{(m-1)!}{(p_{1,j}-1)\ast \prod _{i=1,i\neq j}^{y_{1}}{(p_{1,i})}}}$.
\\For  $\beta =(m)$ ; we obtain ${(m - 1)!} $ distinct permutations on a $m$ symmetric permutation tree, using the above formula.

\subsection[Conjecture ]{Conjecture }
{Any labeled  $(m,2)$  BTU with the first permutation  $p_{1}=I_{m}$  can be represented by a set of ordered labeled partitions on  $m$  elements.}
\\ \textbf{Proof} Since any labeled  $(m,2)$  BTU is isomorphic to  $\Psi (\beta )$ for some \  $\beta \in P_{2}(m)$ , any labeled  $(m,2)$  BTU with the first permutation  $p_{1}=I_{m}$ and second permutation  $p_{2};p_{2}\notin C(p_{1});\beta (p_{1},p_{2});p_{1}=I_{m}$ can be mapped to an ordered labeled partition  $K(\beta ,p_{1})$  on  $m$ in the following manner. Each ordered subset in  $K(\beta ,p_{1})$ consists of  $q_{i}$  elements of distinct labels  $\{l_{i,1},l_{i,2},\ldots ,l_{i,y_{i}}\}$ such that labels  $l_{i,1}$  in  $p_{1}$  and  $l_{i,2}$  in  $p_{2}$  are located at the same depth, \ \ \ labels  $l_{i,2}$  in  $p_{1}$  and  $l_{i,3}$  in  $p_{2}$  are located at the same depth, ..., and finally, \ labels  $l_{i,y_{i}}$  in  $p_{1}$  and  $l_{i,1}$  in  $p_{2}$  are located at the same depth. Thus, 

 $K(\beta ,p_{1})$ has the form  $(l_{1,1},\ldots ,l_{1,y_{1}}),(l_{2,1},\ldots ,l_{2,y_{1}}),$ 

 $\ldots ,(l_{y_{1,}1},\ldots ,l_{y_{1,}y_{1}})$ .

\subsection[Corollary]{Corollary}
{Given an ordered labeled partition on  $m$  represented by  $K(\beta ,p_{1})$ , any circular permutation on ordered subsets yields the same labeled  $(m,2)$  BTU. }
\subsection{Corollary }
{Given an ordered labeled partition on  $m$  represented by  $K(\beta ,p_{1})$ , \ with first permutation  $p_{1}=I_{m}$, $p_{2}$ gets precisely defined resulting in a unique labeled $(m,2)$ BTU.\textit{ }}
\subsection{Conjecture}
{Any labeled $(m,r)$  BTU with the first permutation  $p_{1}=I_{m}$ can be represented by a set of  $r\text{--}1$ ordered labeled partitions on  $m$  elements.}
\\ \textbf{Proof} Let the given \ labeled  $(m,r)$  BTU consists of permutations  $p_{1}=I_{m}$ and $p_{i+1};p_{i+1}\notin C(p_{1},p_{2},\ldots ,p_{i});p_{1}=I_{m};2\le i\le r-1$.Starting with  $p_{1}=I_{m}$, the compatible permutation $p_{2}$ can be represented as an ordered labeled partition  $K_{1}(\beta _{1},p_{1})$ on  $m$ in the following manner. Each ordered subset in  $K_{1}(\beta _{1},p_{1})$ consists of  $q_{1}$ elements of distinct labels  $\{l_{1,1,1},l_{1,1,2},\ldots ,l_{1,1,y_{i}}\}$ such that labels  $l_{1,1,1}$ in  $p_{1}$ and  $l_{1,1,2}$ in  $p_{2}$ are located at the same depth, \ \ \ labels  $l_{1,1,2}$ in  $p_{1}$ and  $l_{1,1,3}$ in  $p_{2}$ are located at the same depth, ..., and finally, \ labels  $l_{1,1,y_{1}}$ in  $p_{1}$ and  $l_{1,1,1}$ in  $p_{2}$ are located at the same depth. 

The compatible permutation $p_{i+1};p_{i+1}\notin C(p_{1},p_{2},\ldots ,p_{i});p_{1}=I_{m};2\le i\le r-1$  can be represented as an ordered labeled partition  $K_{i}(\beta _{i},p_{i})$  on  $m$ in the following manner. Each ordered subset in  $K_{i}(\beta _{i},p_{i})$ consists of  $q_{j};1\le j\le y_{i}$  elements of distinct labels  $\{l_{i,j,1},l_{i,j,2},\ldots ,l_{i,j,y_{i}}\}$ such that labels  $l_{i,j,1}$  in  $p_{i}$  and  $l_{i,j,2}$  in  $p_{i+1}$  are located at the same depth, \ \ \ labels  $l_{i,j,2}$  in  $p_{i}$  and  $l_{i,j,3}$  in  $p_{i+1}$  are located at the same depth, ..., and finally, \ labels  $l_{i,j,y_{i}}$  in  $p_{i}$  and  $l_{i,j,1}$  in  $p_{i+1}$  are located at the same depth.

Thus, any labeled  $(m,r)$  BTU with the first permutation  $p_{1}=I_{m}$ can be represented by a set of  $r\text{--}1$ ordered labeled partitions on  $m$  elements represented by  $\{K_{i}(\beta _{i},p_{i})\};1\le i\le r-1$ . 

Each  $K_{i}(\beta _{i},p_{i})$ has the form  $(l_{i,1,1},l_{i,1,2},\ldots ,l_{i,1,y_{i}}),$ $(l_{i,2,1},l_{i,2,2},\ldots ,l_{i,y_{i}}),\ldots ,$  $(l_{i,y_{i},1},l_{i,y_{i},2},\ldots ,l_{i,y_{i},y_{i}})$ .

\subsection[Conjecture ]{Conjecture\textbf{ }}
{\textup{A set of } $r-1$ \textup{ordered labeled permutations on} $m$ \textup{represented by } $\{K_{i}(\beta _{i},p_{i})\};1\le i\le r-1$ \textup{corresponds to a labeled } $(m,r)$ \textup{ BTU, if the resultant permutations } $\{p_{i+1}\};1\le i\le r-1$ \textup{are compatible with } $\{p_{1},p_{2},\ldots ,p_{i}\};1\le i\le r-1$ \textup{.} }
\\ \textbf{Proof }This follows from the fact that permutations\textbf{ } $\{p_{1},p_{2},\ldots ,p_{r}\}$ correspond to a \ labeled  $(m,r)$  BTU if and only if  $p_{i+1}\notin C(p_{1},p_{2},\ldots ,p_{i})$  for  $1\le i\le r\text{--}1$ . 

\subsection{Conjecture}
If the first permutation  $p_{1}$ of a labeled  $(m,r)$ BTU is specified, and given set of  $r-1$ ordered labeled permutations on  $m$ represented by  $\{K_{i}(\beta _{i},p_{i})\};1\le i\le r-1$ such that the resultant permutations  $\{p_{i+1}\};1\le i\le r-1$ are compatible with  $\{p_{1},p_{2},\ldots ,p_{i}\};1\le i\le r-1$, then the labeled  $(m,r)$ BTU is unique. 

{\bfseries
Proof \textmd{This follows from the fact that permutations}  $\{p_{1},p_{2},\ldots ,p_{r}\}$ \textmd{correspond to a \ labeled } $(m,r)$ \textmd{ BTU if and only if } $p_{i+1}\notin C(p_{1},p_{2},\ldots ,p_{i})$ \textmd{ for } $1\le i\le r\text{--}1$ \textmd{and since the first permutation } $p_{1}$ \textmd{ is specified, we obtain } $p_{2}$ \textmd{ using } $\{K_{(\beta _{1},p_{1})}\}$ \textmd{. Similarly, we obtain } $p_{i+1}$ \textmd{ from } $p_{i}$ \textmd{and } $\{K_{(\beta _{i},p_{i})}\}$ \textmd{for } $1\le i\le r\text{--}1$ \textmd{. We do not }\textmd{separately}\textmd{ need an explicit specification of } $\beta _{i};1\le i\le r-1$ \textmd{since it is already implicitly specified in the definitions of } $\{K_{i}(\beta _{i},p_{i})\};1\le i\le r-1$\textmd{.}}

\subsection[Definition Of Cycle Order ]{Definition Of Cycle Order }
{Given an unordered labeled partition  $U=B_{1}\cup B_{2}\cup \ldots \cup B_{y}$  on  $m$ such that  $B_{i}\cap B_{j}=\Phi $  the empty set for  $i\neq j,1\le i\le y;1\le j\le y$ , and each unordered subset $B_{i}$  contains distinct elements of the set  $\{1,2,\ldots ,m\}$ , a cycle order on a subset  $B_{i}$  is an ordered subset  $C_{i}$ with the same elements such that  $O=C_{1}\cup C_{2}\cup \ldots \cup C_{y}$ is an ordered labeled partition on  $m$ .}
\subsection[Corollary ]{Corollary }
{A defined cycle order on each unordered subset of an unordered \ labeled partition on  $m$ , gives us an ordered labeled partition on  $m$ . }
\subsection[Conjecture ]{Conjecture }
{Any labeled  $(m,r)$  BTU with the first permutation  $p_{1}=I_{m}$ can be represented by a set of  $r\text{--}1$ sets of micro-partitions of  $\beta _{i+1}$  w.r.t.  $\beta _{i}$  for  $1\le i\le r-1$ , mapping from micro-partitions to labels, and defined cycle orders for each partition component. \ }
\\ \textbf{Proof} Any labeled  $(m,r)$  BTU with the first permutation  $p_{1}=I_{m}$ can be represented by a set of  $r\text{--}1$  ordered labeled partitions on $m$ . Each of the  $r\text{--}1$  ordered labeled partitions could be represented by \  $r\text{--}1$  unordered labeled partitions and specific cycle orders for each subset given separately.

Each of the  $r\text{--}1$  unordered labeled partitions could be uniquely specified by \ a set of  $r\text{--}1$ sets of micro-partitions of  $\beta _{i+1}$  w.r.t.  $\beta _{i}$  for  $1\le i\le r-1$ , mapping from micro-partitions to labels. Hence, it follows that any labeled  $(m,r)$  BTU with the first permutation  $p_{1}=I_{m}$ can be represented by a set of  $r\text{--}1$ sets of micro-partitions of  $\beta _{i+1}$  w.r.t.  $\beta _{i}$  for  $1\le i\le r-1$ , mapping from micro-partitions to labels, and defined cycle orders for each partition component. 

\subsection{Corollary}
Given a set of permutations  $p_{1},p_{2},\ldots ,p_{i}$ where  $p_{1}=I_{m}$ and  $p_{i+1}\notin C(p_{1},p_{2},\ldots ,p_{i})$, each permutation satisfying the condition  $\{p_{i+1};\beta _{i}(p_{i+1},p_{i}),p_{i+1}\notin C(p_{1},\ldots ,p_{i})\}$ can be represented by micro-partitions of  $\beta _{i+1}$ w.r.t.  $\beta _{i}$ for  $1\le i\le r-1$,  $i$ unordered labeled partitions corresponding to \ micro-partitions, and finally  $i$ ordered labeled partitions.

\subsection[Micro{}-partitions \ Enumeration Problem]{Micro-partitions \ Enumeration Problem}
Given  $\beta _{i+1},\beta _{i}\in P_{2}(m)$ , the micro-partition enumeration problem refers to enumeration of all possible micro-partitions of  $\beta _{i+1}$ w.r.t.  $\beta _{i}$ .

\subsection[Micro{}-partitions \ Mapping Enumeration Problem]{Micro-partitions \ Mapping Enumeration Problem}
{\bfseries
\textmd{Given} \textmd{micro-partitions of } $\beta _{i+1}$ \textmd{w.r.t. } $\beta _{i}$ \textmd{and } $p_{1},p_{2},\ldots ,p_{i}$ \textmd{ to enumerate all possible }unordered\textmd{ labeled partitions of } $m$ \textmd{ with each subset of labels corresponding to each of the cycles of } $\beta _{i+1}$ \textmd{.}}

\subsection[Conjecture Micro{}-partitions to Permutation Counting ]{Conjecture Micro-partitions to Permutation Counting  $\{p_{3};\beta _{2}(p_{3},p_{2}),p_{3}\notin C(p_{2}, p_{1}=I_{m})\}$ }
Given micro-partitions $x_{1,j,z}$ of  $\beta _{2}$ w.r.t.  $\beta _{1}$ , the number of ways to map the micro-partitions to unordered labeled partitions is given by the following expression  $\prod _{j=1}^{y_{1}}\prod _{z=1}^{y_{2}}\left(\begin{matrix}p_{1,j}-\sum _{k=1}^{z-1}x_{1,j,k}\\x_{1,j,z}\end{matrix}\right)$ .

\textbf{Proof} This follows directly from number of ways to choose  $x_{1,j,z}$ elements from $p_{1,j}-\sum _{k=1}^{z-1}x_{1,j,k}$ elements for  $1\le z\le y_{2}$  and  $1\le j\le y_{1}$ which yields  $\prod _{j=1}^{y_{1}}\prod _{z=1}^{y_{2}}\left(\begin{matrix}p_{1,j}-\sum _{k=1}^{z-1}x_{1,j,k}\\x_{1,j,z}\end{matrix}\right)$ .

\subsection{Cycle Order Enumeration Problem}
Given a set of  $r-1$ unordered labeled partitions of  $m$ , to enumerate all possible distinct $r-1$ ordered labeled partitions, each of which refer a distinct choice of compatible permutation.  $p_{i+1}\notin C(p_{1},p_{2},\ldots ,p_{i})$ w.r.t. \  $p_{1},p_{2},\ldots ,p_{i}$ for  $1\le i\le r\text{--}1$ . Each of enumerated  $r-1$ ordered labeled partitions at each stage have to satisfy the criteria for compatible permutations as far as labels at various depths are concerned in order to correspond to a labeled  $(m,r)$  BTU.

\subsection{Restricted Cycle Order Enumeration Problem}
{\bfseries
\textmd{Given a set of} $r-1$ \textmd{unordered labeled partitions of } $m$ \textmd{, to enumerate all possible distinct} $r-1$ \textmd{ordered labeled partitions, each of which refer a distinct choice of compatible permutation. } $p_{i+1}\notin C(p_{1},p_{2},\ldots ,p_{i})$ \textmd{w.r.t. \ } $p_{1},p_{2},\ldots ,p_{i}$ \textmd{for } $1\le i\le r\text{--}1$ \textmd{. Each of enumerated} $r-1$ \textmd{ordered labeled partitions at each stage have to satisfy the criteria for compatible permutations as far as labels at various depths are concerned in order to correspond to a labeled } $(m,r)$ \textmd{ BTU. For }Restricted Cycle Order Enumeration\textmd{, we impose the following additional constraints}}

\begin{enumerate}
\item 
Without loss of generality, we restrict $p_{1}=I_{m}$.
\item  $p_{2}=\Psi (\beta _{1})$ . The first set of the  $r-1$ ordered labeled partitions on  $m$ corresponding to choices for $p_{2}$  gets fixed due to this.
\item While choosing the second set of of the  $r-1$ ordered labeled partitions of  $m$ corresponding to  $p_{3}\notin C(p_{1},p_{2})$ w.r.t.  $p_{1},p_{2}$ , we choose the first element with the minimum label number from each ordered labeled partitions of  $m$ corresponding to \  $p_{2}\notin C(p_{1})$ w.r.t.  $p_{1}$ . 
\end{enumerate}
We do not lose any non-isomorphic  $(m,r)$  BTU that could be constructed from the set of  $r-1$ unordered labeled partitions of  $m$  in the enumeration process by imposing this constraint. 

\subsection[Micro{}-partitions Isomorphism Theorem for BTUs]{Micro-partitions Isomorphism Theorem for  $(m,r)$ BTUs}
{\bfseries
Enumeration of all non-isomorphic elements in $\Phi (\beta _{1},\beta _{2},\ldots ,\beta _{r-1})$ }

All non-isomorphic  $(m,r)$ BTUs in $\Phi (\beta _{1},\beta _{2},\ldots ,\beta _{r-1})$ where $\beta _{1},\beta _{2},\ldots ,\beta _{r-1}\in P_{2}(m)$ are enumerated at least once by

\begin{enumerate}
\item An exhaustive enumeration of Combinations of Micro-partitions of  $\beta _{i+1}$  w.r.t.  $\beta _{i}$ for  $1\le i\le r\text{--}1$ . 
\item An exhaustive enumeration of sets of  $r\text{--}1$  Unordered labeled partitions for each choice of Combinations of Micro-partitions of  $\beta _{i+1}$  w.r.t.  $\beta _{i}$ for  $1\le i\le r\text{--}1$ enumerated in the previous step.
\item Restricted Cycle Order Enumeration of Sets of  $r\text{--}1$ Ordered labeled partitions that result in compatible permutations corresponding to each Unordered labeled partition enumerated in the previous step.
\item We construct  $p_{2},p_{3},\ldots ,p_{r}$  corresponding to each set of  $r\text{--}1$ Ordered labeled partitions enumerated in the previous step. ( $p_{1}=I_{m}$ )
\end{enumerate}
\textbf{Proof} \ Since any labeled  $(m,r)$  BTU with the first permutation  $p_{1}=I_{m}$ can be represented by a set of  $r\text{--}1$ sets of micro-partitions of  $\beta _{i+1}$  w.r.t.  $\beta _{i}$  for  $1\le i\le r-1$ , mapping from micro-partitions to labels, and defined cycle orders for each partition component, it follows that for each non-isomorphic  $(m,r)$  BTU, there exists at least one set of micro-partitions of  $\beta _{i+1}$  w.r.t.  $\beta _{i}$  for  $1\le i\le r-1$ , one set of \ mapping from micro-partitions to labels, and one set of \ defined cycle orders for each partition component.

Without loss of generality, any non-isomorphic  $(m,r)$  BTU can be mapped to a 

one set of micro-partitions of  $\beta _{i+1}$  w.r.t.  $\beta _{i}$  for  $1\le i\le r-1$ , to one set of \ mapping from micro-partitions to labels using \ \textbf{Restricted Cycle Order Enumeration}, and one set of \ defined cycle orders for each partition component.

\subsection[Corollary Micro{}-partitions Isomorphism for BTUs]{Corollary Micro-partitions Isomorphism for  $(m,3)$ BTUs}
All non-isomorphic $(m,3)$ BTUs in  $\Phi (\beta _{1},\beta _{2})$  can be enumerated by 

\begin{enumerate}
\item An Exhaustive enumeration of Micro-partitions of  $\beta _{2}$  w.r.t.  $\beta _{1}$ .
\item An Exhaustive enumeration of  $2$  Unordered labeled partitions for each choice of Combinations of Micro-partitions of  $\beta _{2}$  w.r.t.  $\beta _{1}$  enumerated in the previous step.
\item Restricted Cycle Order Enumeration of Set of  $2$ Ordered labeled partitions that result in compatible permutations corresponding to each set of  $2$ Unordered labeled partitions enumerated in the previous step.
\item We construct  $p_{2},p_{3}$  corresponding to the set of  $2$ Ordered labeled partitions enumerated in the previous step. ( $p_{1}=I_{m}$ ).
\end{enumerate}
\subsection[Search Problem for BTU with best girth in ]{Search Problem for BTU with best girth in  $\Phi (\beta _{1},\beta _{2},\ldots ,\beta _{r-1})$ }
{\bfseries
\textmd{Given}  $\beta _{1},\beta _{2},\ldots ,\beta _{r-1}\in P_{2}(m)$ , \textmd{we} \textmd{exhaustively enumerate}}

\begin{enumerate}
\item Combinations of Micro-partitions of  $\beta _{i+1}$  w.r.t.  $\beta _{i}$ for  $1\le i\le r\text{--}1$ . 
\item A set of  $r\text{--}1$  Unordered labeled partitions for each choice of Combinations of Micro-partitions of  $\beta _{i+1}$  w.r.t.  $\beta _{i}$ for  $1\le i\le r\text{--}1$ enumerated in the previous step.
\item Sets of  $r\text{--}1$ Ordered labeled partitions that result in compatible permutations corresponding to each set of  $r\text{--}1$ Unordered labeled partition enumerated in the previous step.
\item We construct  $p_{2},p_{3},\ldots ,p_{r}$  corresponding to the set of  $r\text{--}1$ Ordered labeled partitions enumerated in the previous step. ( $p_{1}=I_{m}$ ).
\item We evaluate the girth for each of the enumerated/constructed  $(m,r)$  BTUs.
\item We choose the BTU with the best girth at the end of this process.
\end{enumerate}
\section{Direct Construction}
\subsection[Generalized Cycle Traversal for a permutation representation of a  BTU]{Generalized Cycle Traversal for a permutation representation of a  $(m,r)$  BTU}
If labels $l_{1}$  and  $l_{2}$  occur at the same depth, labels $l_{2}$  and  $l_{3}$  occur at the same depth,  $\ldots $ , and finally labels $l_{x}$ and  $l_{1}$  occur at the same depth, in the permutation representation of a \ labeled  $(m,r)$ BTU, then there exists a cycle connecting the labels  $l_{1},l_{2,}\ldots ,l_{x}$ .

\subsection[Known Cycle Conjecture for a  BTU]{Known Cycle Conjecture for a  $(m,2)$  BTU}
{\textup{The cycle lengths of a} $(m,2)$ \textup{ BTU that is isomorphic to } $\Psi (\beta )$ \textup{for some } $\beta \in P_{2}(m)$ \textup{given by } $\sum _{i=1}^{y}q_{i}=m$ \textup{ are } $\{2\ast q_{i}\};1\le i\le y$ \textup{ .}}
\\  \textbf{Proof} A $(m,2)$  BTU that is isomorphic to  $\Psi (\beta )$ has no other cycles other than that of \  $\beta \in P_{2}(m)$ given by  $\sum _{i=1}^{y}q_{i}=m$ . The cycle length for a partition component  $q_{i}$  is  $2\ast q_{i}$ . Hence, it follows that the cycle lengths are are  $\{2\ast q_{i}\};1\le i\le y$  .

\subsection[Maximum possible girth of a  BTU]{Maximum possible girth of a  $(m,2)$  BTU}
{The maximum possible girth of a  $(m,2)$  BTU is $2\ast m$ . }
\\ \textbf{Proof} This directly follows when we consider that every  $(m,2)$ BTU can be mapped to  $\Psi (\beta )$ where  $\beta \in P_{2}(m)$ . It is clear that girth of a  $(m,2)$ BTU is  $2\ast \mathit{min}(q_{i});1\le i\le y$ where $\sum _{i=1}^{y}q_{i}=m$ represents $\beta \in P_{2}(m)$ . Hence, it follows that the maximum possible girth of a $(m,2)$  BTU is  $2\ast m$ . 

\subsection[Upper Bounds Known partition component upper bound Conjecture ]{Upper Bounds Known partition component upper bound Conjecture }
If $u=\mathit{min}(q_{i,j};1\le j\le y_{i};1\le i\le r\text{--}1)$ where each partition $\beta _{i}$  is given by  $\sum _{j=1}^{y_{i}}q_{i,j}=m$ for  $\beta _{1},\beta _{2},\ldots ,\beta _{r-1}\in P_{2}(m)$ where each  $u,q_{i,j}\in \mathbb{N}$ , then the maximum possible girth of all $(m,r)$  BTUs in  $\Phi (\beta _{1},\beta _{2},\ldots ,\beta _{r-1})$ is less than or equal to $2\ast u$ . This is an upper bound on the possible possible girth. 

\textbf{Proof} This follows directly from the fact that there can be smaller cycles caused due to interactions between the partitions  $\beta _{1},\beta _{2},\ldots ,\beta _{r-1}\in P_{2}(m)$ and the maximum girth of all $(m,r)$  BTUs in  $\Phi (\beta _{1},\beta _{2},\ldots ,\beta _{r-1})$ is less than or equal to $2\ast u$ , with strict equality when  $r=2$.

\section[\ Micro{}-partition cycles]{\ Micro-partition cycles}
For a  $(m,r)$ BTU $\Phi (\beta _{1},\beta _{2},\ldots ,\beta _{r-1})$ where\textbf{ } $\beta _{1},\beta _{2},\ldots ,\beta _{r-1}\in P_{2}(m)$ given by  $\{p_{1},p_{2},\ldots ,p_{r}\};p_{i+1}\notin C(p_{1},p_{2},\ldots ,p_{i})$ for  $1\le i\le r\text{--}1$. If each  $\beta _{i}$ refers to  $\sum _{j=1}^{y_{i}}q_{i,j}=m$ for  $1\le i\le r-1$, the micro-partition cycles are the cycles caused by interaction between the partition components of  $\beta _{u}$ and  $\beta _{v}$ where  $u\neq v;1\le u\le r-1;1\le v\le r-1$ are the cycles caused due to interactions between the known cycles of  $\beta _{u}$ and  $\beta _{v}$ namely \  $\{q_{u,1},q_{u,2},\ldots ,q_{u,y_{u}}\}$ and  $\{q_{v,1},q_{v,2},\ldots ,q_{v,y_{v}}\}$ and corresponding micro-partitions  $\{x_{u,v,c,d}\};1\le c\le y_{u};1\le d\le y_{v}$ and  $\{x_{v,u,d,c}\};1\le c\le y_{u};1\le d\le y_{v}$.

\subsection{When do micro-partition cycles arise?}
When $x_{u,v,c,d}\neq 1$ and  $x_{u,v,c,d}\neq 0$ for some $\{c,d\}$ where  $1\le c\le y_{u};1\le d\le y_{v}$ , we have more than one point from partition component  $q_{u,c}$ is used for creating partition component $q_{v,d}$ , then we have an additional cycle referred to as micro-partition cycle.

\subsection[\ \ Conjecture \ for Length of micro{}-partition cycle]{\ \ Conjecture \ for Length of micro-partition cycle}
For a $(m,r)$ BTU in  $\Phi (\beta _{1},\beta _{2},\ldots ,\beta _{r-1})$ where  $\beta _{1},\beta _{2},\ldots ,\beta _{r-1}\in P_{2}(m)$ , \ the maximum possible length of micro-partition cycle is $\mathit{min}\{2\ast (q_{u,j}/x_{u,v,j,z}+t_{u}-1)\};$   $x_{u,v,j,z}\ge 2;1\le j<y_{u};1\le z\le y_{v}$ where $\beta _{i}$  refers to  $\sum _{j=1}^{y_{i}}q_{i,j}=m$ for  $1\le i\le r-1;i\in \mathbb{N}$ , where Generalized Micro-partition between $\beta _{u},\beta _{v}\in P_{2}(m);u\neq v;$  $1\le u\le r\text{--}1;1\le v\le r\text{--}1$ :  $x_{u,v,j,z}$  where $1\le j\le y_{u};1\le z\le y_{v}$ . \  $t_{u}$ which is the number of partition components of  $\beta _{u}$ connected by the micro-partition cycle ,  $1\le t_{u}\le y_{u}$ .

{\textbf{\textup{Proof}}\textup{ If} $x_{u,v,c,d}\neq 1$ \textup{and } $x_{u,v,c,d}\neq 0$ \textup{for some} $\{c,d\}$ \textup{where } $1\le c\le y_{u};1\le d\le y_{v}$ \textup{, we have more than one point from partition component } $q_{u,c}$ \textup{ is used for creating partition component} $q_{v,d}$ \textup{, then we have a micro-partition cycle. \ \ Since the number of partition components of } $\beta _{u}$ \textup{connected by the micro-partition cycle \ is } $t_{u}$ \textup{, where} $1\le t_{u}\le y_{u}$ \textup{, we have a smaller cycle which can take the maximum value } $\{2\ast (q_{u,j}/x_{u,v,c,d}+t_{u}-1)\}$ \textup{, by choosing the } $x_{u,v,c,d}$ \textup{points appropriately. Hence , the maximum possible length of micro-partition cycle is} $\mathit{min}\{2\ast (q_{u,j}/x_{u,v,j,z}+t_{u}-1)\};$\textup{ } $x_{u,v,j,z}\ge 2;1\le j<y_{u};1\le z\le y_{v}$\textup{.}}
\subsection[Corollary for BTU]{Corollary for  $(m,3)$ BTU}
For a $(m,3)$ BTU in  $\Phi (\beta _{1},\beta _{2})$ where  $\beta _{1},\beta _{2}\in P_{2}(m)$ , if micro-partition cycles exist, the minimum possible length of micro-partition cycle is $\mathit{min}\{2\ast (q_{1,j}/x_{1,j,z}+t_{1}-1)\}$ where the micro-partitions  $x_{1,j,z}\ge 2;1\le j<y_{1};1\le z\le y_{2}$ where $\beta _{i}$  refers to $\sum _{j=1}^{y_{i}}q_{i,j}=m$ . \  $t_{1}$ which is the number of partition components of  $\beta _{1}$ connected by the micro-partition cycle ,  $\mathit{max}(t_{1})=y_{1}$  .

\subsection[All cycles caused ]{All cycles caused }
Given a labeled  $(m,r)$ BTU with compatible permutations  $\{p_{1},p_{2},\ldots ,p_{r}\}$ in  $\Phi (\beta _{1},\beta _{2},\ldots ,\beta _{r-1})$ where each  $\beta _{i}\in P_{2}(m)$
refers to $\sum _{j=1}^{y_{i}}q_{i,j}=m$ for  $1\le i\le r\text{--}1$ we categorize its cycles in the following manner
\begin{enumerate}
\item Known cycles : These cycles refer to the cycles  $\{q_{i,j}\}$ for  $1\le j\le y_{i}$ and  $1\le i\le r\text{--}1$ corresponding to  $\beta _{1},\beta _{2},\ldots ,\beta _{r-1}$ .We have  $r\ast (r\text{--}1)/2$ partitions in total arising from all possible combinations of  $2$  permutations from the set of  $r$  permutations, out of which  $r\text{--}1$ partitions are considered for known cycles. 
\item Cycles due to other partitions: We consider generalized partitions  $\alpha _{u,v}\in P_{2}(m)$ where  $\left|(u-v)\right|\neq 0$, $\left|(u-v)\right|\neq 1$ and  $1\le u\le r;1\le v\le r$. The number of partitions considered here are 

 $r\ast (r\text{--}1)/2\text{--}(r\text{--}1)=r^{2}/2\text{--}r/2+1$ .
\item Micro-partition cycles if they arise due to interactions between the combinations of  $r\ast (r\text{--}1)/2$ permutations.
\item Hidden cycles caused due to interaction of all the above cycles. 
\end{enumerate}
\subsection[Strategy for girth maximization]{Strategy for girth maximization}
In order to construct a  $(m,r)$ BTU with maximum girth, we choose

\begin{enumerate}
\item Partitions  $\beta _{1},\beta _{2},\ldots ,\beta _{r-1}\in P_{2}(m)$ such that the known cycles are maximized.
\item By maximizing the length of the micro-partition cycle, we obtain optimal parameters for  $\beta _{1},\beta _{2},\ldots ,\beta _{r-1}$ .
\item We search for permutations such that the cycles due to other partitions and hidden cycles \ caused due to interaction of all the above cycles is maximized.
\end{enumerate}
\subsection[Self Evident Fact About the Girth of a member of]{Self Evident Fact About the Girth of a member of $\Phi (\beta _{1},\beta _{2},\ldots ,\beta _{r-1})$ }
If  $2\ast v$ is the length of the minimum micro-partition cycle, and \  $u\in \mathbb{N}$ is the smallest partition component among given $\beta _{1},\beta _{2},\ldots ,\beta _{r-1}\in P_{2}(m)$ i.e.,  $u=\mathit{min}(q_{i,j};1\le j\le y_{i};1\le i\le r\text{--}1)$ where each  $\beta _{i}$ is given by $\sum _{j=1}^{y_{i}}q_{i,j}=m$, and  $2\ast w\in \mathbb{N}$ is the the length of the smallest cycle caused due to interactions between the micro-partition cycles and cycles due to  $\beta _{1},\beta _{2},\ldots ,\beta _{r-1}$, the girth of a member of $\Phi (\beta _{1},\beta _{2},\ldots ,\beta _{r-1})$ is $2\ast \mathit{min}(u,v,w)$ or  $2\ast \mathit{min}(u,w)$ if there are no micro-partition cycles or  $2\ast u$ for $r=2$, in which case the only cycles that arise due to one element of  $P_{2}(m)$.

\subsection[\  BTU puncturing Conjecture]{\textup{\ } $(k,2)$ \textup{ BTU puncturing }Conjecture}
If a $0$ element in a  $(k,2)$ BTU with cycle length  $2\ast k$ is changed to  $1$, then length of new minimum cycle within the \ sub-block  $l\in \mathbb{N}$ satisfies $4\le l\le k$ if  $k$ is an even positive integer and $4\le l\le k+1$ if  $k$ is an odd positive integer.

{\textbf{\textup{Proof}}\textup{ \ Let us map the} $(k,2)$ \textup{BTU to a labeled directed graph with k vertices such that each vertex } $i;1\le i\le k$\textup{ is connected to vertex} $i+1$ \textup{mod} $k$\textup{. \ If the distance between two vertexes is defined as the lenght of shortest traversals in the same direction of the directed edges, it is clear that the maximum distance measured in terms of number of directed traversals from one vertex to the next, between two vertexes on this labeled directed graph is } $k/2$ \textup{for even positive integers } $k$\textup{ and \ } $(k+1)/2$ \textup{for odd positive integers } $k$\textup{, and the \ minimum distance measured in terms of number of directed traversals from one vertex to the next, between two vertexes on this labeled directed graph is} $1$\textup{.}}
If a directed edge is connected between two vertexes of minimum distance of  $1$ , this leads to a minimum cycle length of  $4$  on the matrix representation. If a directed edge is connected between two vertexes of maximum distance $k/2$ for even positive integers  $k$  and \  $(k+1)/2$ for odd positive integers  $k$ , we get a cycle length of  $k$ for even positive integers  $k$  and \  $(k+1)$ for odd positive integers  $k$ in the equivalent matrix representation.

Hence, the length of new minimum cycle within the \ sub-block  $l\in \mathbb{N}$ satisfies $4\le l\le k$ if  $k$  is an even positive integer and $4\le l\le k+1$ if  $k$  is an odd positive integer.

\subsection[Three {}-one Conjecture]{Three -one Conjecture}
Let us consider a  $(2\ast k,2)$ BTU constructed with  $p_{1}=I_{2\ast k}$ and  $p_{2}$ as per $\Psi ((k,k))$, and if we have to additionally convert three $0$ s in this BTU to  $1$ s, the girth is strictly less than  $2\ast k$ .

\textbf{Proof} Girth of a $(2\ast k,2)$ BTU constructed with  $p_{1}=I_{2\ast k}$ and  $p_{2}$ as per $\Psi ((k,k))$ is  $2\ast k$. Let us denote the $(2\ast k,2)$ BTU consisting of sub-matrices  $B_{1}$ and  $B_{2}$ each of which are  $k \times k$ matrices that represent a \ constituent  $(k,2)$  BTU, and two $k \times k$  matrices referred to as  $\mathit{CB}(1,2)$ that shares its rows with $B_{1}$ and columns with $B_{2}$ and  $\mathit{CB}(2,1)$ that shares its rows with $B_{2}$ and columns with $B_{1}$.

Let us consider different cases for placement of the \ three  $1$ s.

Case 1: If a $1$ is placed inside either of the constituent $(k,2)$ BTUs, by the previous theorem, the girth reduces to  $k+1$ if  $k$ is odd, and  $k$ if  $k$ is even. 

Case 2: Three  $1$ s in  $\mathit{CB}(1,2)$ 

Let the positions of the three $1$ s in  $\mathit{CB}(1,2)$ be  $(x_{1},y_{1}),(x_{2,}y_{2}),(x_{3,}y_{3})$ such that  $1\le x_{i}\le k;1\le y_{i}\le k$ for  $1\le i\le 3$ and $x_{1}\neq x_{2};x_{2}\neq x_{3};x_{3}\neq x_{1};y_{1}\neq y_{2};y_{2}\neq y_{3};y_{3}\neq y_{1}$. \ Let us define  $h(w_{1},w_{2},k)=\mathit{min}\{\left|(w_{1}\text{--}w_{2})\right|,k-\left|(w_{1}\text{--}w_{2})\right|\}$

We can verify that traversals lengths between any two of the three points through  $B_{1}$ are  $2\ast h(x_{1},x_{2},k)+1$,  $2\ast h(x_{2},x_{3},k)+1$ and  $2\ast h(x_{3},x_{1},k)+1$.

We can verify that traversals lengths between any two of the three points through  $B_{2}$ are  $2\ast h(y_{1},y_{2},k)+1$ ,  $2\ast h(y_{2},y_{3},k)+1$ and  $2\ast h(y_{3},y_{1},k)+1$ . The corresponding cycle lengths are  $2\ast h(x_{1},x_{2},k)+2\ast h(y_{1},y_{2},k)+2$ , $2\ast h(x_{2},x_{3},k)+2\ast h(y_{2},y_{3},k)+2$  and $2\ast h(x_{3},x_{1},k)+2\ast h(y_{3},y_{1},k)+2$  .

If possible let the length of the minimum cycle be greater than or equal to $2\ast k$, which implies that 

\  $h(x_{1},x_{2},k)+h(y_{1},y_{2},k)\ge k-1$ , $h(x_{2},x_{3},k)+h(y_{2},y_{3},k)\ge k-1$  and $h(x_{3},x_{1},k)+h(y_{3},y_{1},k)\ge k-1$  . This gives rise to a contradiction since the upper bound on maximum attainable value of \ 

 $\mathit{min}(\left|(x_{1}-x_{2})\right|,\left|(x_{2}-x_{3})\right|,\left|(x_{3}-x_{1})\right|)$ is $k/3$ and similarly upper bound on maximum attainable value of \ 

 $\mathit{min}(\left|(y_{1}-y_{2})\right|,\left|(y_{2}-y_{3})\right|,\left|(y_{3}-y_{1})\right|)$ is $k/3$ .

Hence. The maximum attainable girth when three  $1$ s are placed in  $\mathit{CB}(1,2)$ is strictly less than  $2\ast k$. 

\ Case 3: Three  $1$ s in  $\mathit{CB}(2,1)$ 

By repeating the argument for three  $1$ s in  $\mathit{CB}(1,2)$ we can show that the maximum attainable girth when three  $1$  s are placed in  $\mathit{CB}(2,1)$  is strictly less than  $2\ast k$ . 

\ Case 4: Two  $1$ s placed in  $\mathit{CB}(1,2)$  and one $1$ placed in  $\mathit{CB}(2,1)$ 

Maximum value of traversal length from one point to another through $B_{1}$ is  $k/3$. Maximum value of traversal length from one point to another through $B_{2}$ is  $k/3$ . Hence. The maximum attainable girth when two  $1$ s placed in  $\mathit{CB}(1,2)$  and one $1$ placed in  $\mathit{CB}(2,1)$ is strictly less than  $2\ast k$ . 

Case 5: Two  $1$ s placed in  $\mathit{CB}(2,1)$  and one $1$ placed in  $\mathit{CB}(1,2)$ 

By repeating the argument for two  $1$ s placed in  $\mathit{CB}(1,2)$  and one $1$ placed in  $\mathit{CB}(2,1)$ we can show that the maximum attainable girth when two  $1$ s placed in  $\mathit{CB}(2,1)$  and one $1$ placed in  $\mathit{CB}(1,2)$  is strictly less than  $2\ast k$ . 

Hence, \ given a $(2\ast k,2)$ BTU constructed with  $p_{1}=I_{2\ast k}$  and  $p_{2}$  as per $\Psi ((k,k))$ , and if we have to additionally convert three $0$ s in this BTU to  $1$ s, the girth is strictly less than  $2\ast k$ .

\subsection[Upper bound on the maximum attainable girth for the case when no micro{}-partition cycles arises for a  BTU]{Upper bound on the maximum attainable girth for the case when no micro-partition cycles arises for a  $(k^{2},3)$ BTU}
No micro-partition cycles arise when $\beta _{1},\beta _{2}\in P_{2}(m)$ both correspond to the partition $\sum _{j=1}^{k}{k}=k^{2}$. Let us construct a labeled $(k^{2},2)$ BTU with $p_{1}=I_{k^{2}}$ and $p_{2}\in S_{k^{2}}$ as per  $\Psi (\beta _{1})$. \ Let us consider the labeled $(k^{2},2)$ BTU as consisting of \  $k$  constituent  $(k,2)$ BTUs which we refer to as sub-blocks numbers from  $\{1,2,\ldots ,k\}$ and $k^{2}\text{--}k$ cross-blocks  $\mathit{CB}(i,j)$ where  $1\le i\le k;1\le j\le k;i\neq j$. Each cross-block  $\mathit{CB}(i,j)$ shares its rows with sub-block  $i$ and shares its columns with sub-block  $j$ .

Let $p_{3}\in S_{k^{2}}$ be the permutation that maximizes the girth among all possible  $(k^{2},3)$ BTUs in  $\Psi (\beta _{1},\beta _{2})$. 

There must exist at least one cross-block in each cross-block row that has two  $1$ s from  $p_{3}$. There must exist at least one cross-block in each cross-block column that has two  $1$ s from  $p_{3}$ .

Hence, there must exist $u,v$ such that  $u\neq v$ and  $1\le u\le k;1\le v\le k$ such that  $\mathit{CB}(u,v)$ and  $\mathit{CB}(v,u)$ have two  $1$ s and one  $1$ respectively.

If we consider a  $2\ast k \times 2\ast k$ matrix consisting of a $(k,2)$ BTU  $B_{1}$ and  $\mathit{CB}(u,v)$ in the same rows, and  $(k,2)$ BTU  $B_{2}$ and  $\mathit{CB}(u,v)$ in the same columns and consequently,  $B_{1}$ and  $\mathit{CB}(v,u)$ share the same columns and  $B_{2}$ and  $\mathit{CB}(v,u)$ share the same rows.

By using the previously proved result, girth is strictly less than  $2\ast k$ .

Hence, the maximum attainable girth \ of  $(k^{2},3)$ BTU with no micro-partition cycles is strictly less than  $2\ast k$ .

\subsection{Conjecture}
The maximum attainable girth for a  $(k^{2},3)$ BTU for the case when micro-partition cycles arise is greater than the maximum attainable girth for a  $(k^{2},3)$ BTU for the case when no micro-partition cycles arise.

\textbf{Proof} For the case where no micro-partition cycles arise, let  $p_{3}\in S_{k^{2}}$ maximize the girth among all possible  $(k^{2},3)$ BTUs in  $\Psi (\beta ,\beta )$ where  $\beta \in P_{2}(m)$  corresponds to the partition  $\sum _{j=1}^{k}{k}=k^{2}$ with $p_{1}=I_{k^{2}}$ and $p_{2}\in S_{k^{2}}$ as per  $\Psi (\beta )$ . Let us consider the labeled $(k^{2},2)$ BTU as consisting of \  $k$  constituent  $(k,2)$ BTUs which we refer to as sub-blocks numbers from  $\{1,2,\ldots ,k\}$ and $k^{2}\text{--}k$ cross-blocks  $\mathit{CB}(i,j)$ where  $1\le i\le k;1\le j\le k;i\neq j$ . Each cross-block  $\mathit{CB}(i,j)$ shares its rows with sub-block  $i$ and shares its columns with sub-block  $j$ .

Let us choose the following $\beta _{1},\beta _{2}\in P_{2}(m)$ that correspond to  $\sum _{j=1}^{k}k=k^{2}$ and $\sum _{j=1}^{1}k^{2}=k^{2}$ respectively. Without loss generality,  $p_{3}$ is such that $\mathit{CB}(1,k-1),\mathit{CB}(2,1),\ldots ,\mathit{CB}(k,k-2)$ and  $\mathit{CB}(k-1,1),\mathit{CB}(1,2),\ldots ,\mathit{CB}(k-2,k)$ such that each pair of cross-blocks  $\{\mathit{CB}(1,k\text{--}1),\mathit{CB}(k\text{--}1,1)\}$,  $\{\mathit{CB}(2,1),\mathit{CB}(1,2)\}$ ,  $\ldots $,

 $\{\mathit{CB}(k,k\text{--}2),\mathit{CB}(k\text{--}2,k)\}$ have exactly two  $1$ s between the two of them \ such that each $1$ with coordinates  $(x,y);1\le x\le k^{2};1\le y\le k^{2}$

satisfies the constraint  $\left|(x\text{--}y)\right|\ge k$.

We now replace  $p_{2}$  with  $q_{2}\in S_{k^{2}}$  as per  $\Psi (\beta _{2})$ .

Now, the cycle length due to  $\{\mathit{CB}(1,k\text{--}1),\mathit{CB}(k\text{--}1,1)\}$ and sub-blocks  $1$ and  $k$ is now  $2\ast k$ since the function for traversal length between two points  $(x_{1,}y_{1})$ and  $(x_{2,}y_{2})$ 

is now  $2\ast \left|(x_{1}\text{--}x_{2})\right|+1$ and  $2\ast \left|(y_{1}\text{--}y_{2})\right|+1$ instead of  $2\ast \mathit{min}\{\left|(x_{1}\text{--}x_{2})\right|,k-\left|(x_{1}\text{--}x_{2})\right|\}+1$ and 

 $2\ast \mathit{min}\{\left|(y_{1}\text{--}y_{2})\right|,k-\left|(y_{1}\text{--}y_{2})\right|\}+1$ .

The same is true for  $\{\mathit{CB}(2,1),\mathit{CB}(1,2)\}$ ,  $\ldots $ ,

 $\{\mathit{CB}(k,k\text{--}2),\mathit{CB}(k\text{--}2,k)\}$ .

Now, let us examine the situation that leads to the constraint on maximum attainable minimum cycle length when no micro-partitions cycles arise, and see that the maximum attainable minimum cycle length is better for when micro-partitions cycles arise.

Let us consider $\mathit{CB}(u,v)$ and  $\mathit{CB}(v,u)$ with 

with three $1$ s between both the cross-blocks in the pair. 

Case 1: Three  $1$ in  $\mathit{CB}(u,v)$  zero  $1$ s in  $\mathit{CB}(v,u)$ .

Let the positions of the three $1$ s in $\mathit{CB}(u,v)$ be  $(x_{1},y_{1}),(x_{2,}y_{2}),(x_{3,}y_{3})$ such that  $1\le x_{i}\le k;1\le y_{i}\le k$  for  $1\le i\le 3$ and $x_{1}\neq x_{2};x_{2}\neq x_{3};x_{3}\neq x_{1};y_{1}\neq y_{2};y_{2}\neq y_{3};y_{3}\neq y_{1}$ . \ Let us define  $h_{2}(w_{1},w_{2})=\left|(w_{1}\text{--}w_{2})\right|$ 

We can verify that traversals lengths between any two of the three points through  $B_{1}$ are  $2\ast h_{2}(x_{1},x_{2})+1$ ,  $2\ast h_{2}(x_{2},x_{3})+1$ and  $2\ast h_{2}(x_{3},x_{1})+1$ .

We can verify that traversals lengths between any two of the three points through  $B_{2}$ are \  $2\ast h_{2}(y_{1},y_{2})+1$ ,  $2\ast h_{2}(y_{2},y_{3})+1$ and  $2\ast h_{2}(y_{3},y_{1})+1$ .The corresponding cycle lengths are  $2\ast h_{1}(x_{1},x_{2})+2\ast h_{2}(y_{1},y_{2})+2$ , $2\ast h_{2}(x_{2},x_{3})+2\ast h_{2}(y_{2},y_{3})+2$  and $2\ast h_{2}(x_{3},x_{1})+2\ast h_{2}(y_{3},y_{1})+2$  .

Thus, the length of the minimum cycle is increased compared to the previous traversal length  $2\ast h(y_{3},y_{1},k)+1$  where  $h(y_{3},y_{1},k)=\mathit{min}\{\left|(y_{1}\text{--}y_{2})\right|,k-\left|(y_{1}\text{--}y_{2})\right|\}$.

Case 2: We can similarly show that the one Cross-block has two 1s and other in the pair has one 1s, the length of the minimum cycle is increased for the case when micro-partition cycles arise.

Similarly, we can also show that for \ the case when

 $\mathit{CB}(u,v)$  and  $\mathit{CB}(v,u)$  with 

with two $1$ s between both the cross-blocks in the pair, has the length of the maximum cycle increased.

Given any points corresponding to  $p_{3}$, we can show that the traversal length increases for the considered case when micro-partition cycles arise. 

Hence, the maximum attainable girth for a  $(k^{2},3)$ BTU for the case when micro-partition cycles arise is greater than the maximum attainable girth for a  $(k^{2},3)$ BTU for the case when no micro-partition cycles arise.


\subsection[Girth maximum Conjecture forBTU]{Girth maximum Conjecture for $(k^{2},3)$ \textup{BTU}}
{If  $k\in \mathbb{N};k>3$ there exists a $(k^{2},3)$ BTU in $\Phi (\beta _{1},\beta _{2})$ with maximum girth among all  $(k^{2},3)$ BTUs where each $\beta _{i}\in P_{2}(k^{2})$  refers to  $\sum _{j=1}^{k^{3-1-i}}\{k^{i}\}=k^{2}$ for $1\le i\le 2$ .}
\textbf{Proof} Let $\beta _{1},\beta _{2}\in P_{2}(b\ast k^{2})$ be of the form  $\sum _{j=1}^{y_{i}}q_{i,j}=k^{2}$ for  $1\le i\le 2$. Since we have proved that the case where no micro-partition cycles arise produces lesser minimum cycle length than the case where \ micro-partition cycles arise, let us maximize the length of micro-partition cycles. 

Micro-partition cycles and cycles due to $\beta _{1},\beta _{2}$ can produce other interacting cycles that are of smaller length. Hence, let us maximize the length of the \ minimum micro-partition cycle. Since the length of the minimum micro-partition cycle would be less than or equal to $\mathit{min}\{2\ast (q_{1,j}/x_{1,j,z}+t_{1}-1)\}$, for the case for micro-partition cycles maximized, we obtain $q_{2,1}=k^{2};y_{1}=k;q_{1,j}=k;y_{2}=1$ and micro-partitions  $x_{1,2,j,1}=k$ for $1\le j\le k$ , $x_{2,1,1,j}=k$ for $1\le j\le k$ 

Micro-partition cycles are 

 $2\ast (k^{2}/k+1\text{--}1)=2\ast k$ and

 $2\ast (k/k+k\text{--}1)=2\ast k$ .

Starting with $p_{1}=I_{k^{2}}$  and $p_{2}\in S_{k^{2}}$  as per  $\Psi (\beta _{1})$ , we can choose $p_{3}\in S_{k^{2}}$ by considering the interactions between the micro-partition cycles and known cycles , i.e., one cycle of length  $2\ast k^{2}$ and  $k$ cycles of length  $2\ast k$, we construct a girth maximum  $(k^{2,}3)$ BTU. Given $k\in \mathbb{N}$, thus $\exists $ a BTU with maximum girth among all $(k^{2},3)$ BTUs in  $\Phi (\beta _{1},\beta _{2})$ where \  $\beta _{1},\beta _{2}\in P_{2}(m)$ correspond to  $\sum _{j=1}^{k}k=k^{2}$  and $\sum _{j=1}^{1}k^{2}=k^{2}$ respectively.

\subsection[Conjecture]{Conjecture}
If  $k\in \mathbb{N};k>3$ and $b\in \mathbb{N};b^{2}<k$ there exists a \ \ \  $(b\ast k^{2},3)$ BTU in $\Phi (\beta _{1},\beta _{2})$ with maximum girth among all  $(b\ast k^{2},3)$ BTUs where each $\beta _{i}\in P_{2}(b\ast k^{2})$  refers to  $\sum _{j=1}^{k^{3-1-i}}\{b\ast k^{i}\}=b\ast k^{2}$ for  $1\le i\le 2$ .

\subsection{Notation for scaling of a partition }
Scaling of a partition  $\alpha \in P_{2}(m)$ which refers to  $\sum _{j=1}^{y}q_{j}=m$ by  $k$ is denoted by $k\ast \alpha \in P_{2}(k\ast m)$ which refers to the partition $\sum _{j=1}^{k\ast y}q_{j}=k\ast m$.

\subsection[Conjecture for girth maximumBTU]{Conjecture for girth maximum $(b\ast k^{r-1},r)$ BTU}
{\textup{If } $k,r\in \mathbb{N};k>r$ \textup{and } $b=\prod _{i=1}^{r-1}b_{i}$ \textup{such that } $b_{i}\in \mathbb{N}$ \textup{for} $1\le i\le r-1$ \textup{satisfy \ \ \ } $b_{1}\le b_{2}\le \ldots \le b_{r-1}<k$ \textup{and } $b_{1}=1$\textup{, t}\textup{here exists a } $(b\ast k^{r-1},r)$ \textup{BTU in} $\Phi (\beta _{1},\beta _{2},\ldots ,\beta _{r-1})$ \textup{with maximum girth among all} $(b\ast k^{r-1},r)$ \textup{BTUs }\textup{where each } $\beta _{i}\in P_{2}(b\ast k^{r-1})$ \textup{refers to } $\sum _{j=1}^{k^{r-1-i}}\{b\ast k^{i}\}=b\ast k^{r-1}$ \textup{for } $1\le i\le r-1$\textup{.}}
\\ \textbf{Proof} We prove the above statement using the principle of mathematical induction. For  $r=3$, with $p_{1}=I_{b\ast k^{2}}$ and  $p_{2}$ as per  $\Psi (\beta _{1})$ where $\beta _{1}\in P_{2}(b\ast k^{2})$ refers to $\sum _{j=1}^{k}\{b\ast k^{1}\}=b\ast k^{2}$, and $\beta _{2}\in P_{2}(b\ast k^{2})$ refers to  $\sum _{j=1}^{1}\{b\ast k^{2}\}=b\ast k^{2}$ .

Since $(b\ast k^{1},2)$ BTU with $\sum _{j=1}^{1}\{b\ast k^{1}\}=b\ast k^{1}$ has maximum girth among all $(b\ast k^{1},2)$ BTUs, we choose  $p_{3}\in S_{b\ast k^{2}}$ so that partition between  $p_{2}$ and  $p_{3}$ \ is $\sum _{j=1}^{1}\{b\ast k^{2}\}=b\ast k^{2}$ , such that we get maximum girth, clearly there exists a BTU with maximum girth in  $\Phi (\beta _{1},\beta _{2})$ . Hence the statement is proven for  $r=3$.

Let us assume that the statement is true for $r=l$, we need to prove that it is also true for $r=l+1$ .

We assume that there exists  $(b\ast k^{l-1},l)$  BTU with maximum girth in $\Phi (\lambda _{1},\lambda _{2},\ldots ,\lambda _{l-1})$ where each  $\lambda _{i}\in P_{2}(b\ast k^{l-1})$ refers to $\sum _{j=1}^{k^{l-1-i}}\{b\ast k^{i}\}=b\ast k^{l-1}$ for  $1\le i\le l-1$ . 

We need to prove that there exists $(b\ast k^{l},l+1)$  BTU with maximum girth in $\Phi (\beta _{1},\beta _{2},\ldots ,\beta _{l})$  where each  $\beta _{i}\in P_{2}(b\ast k^{l})$ refers to $\sum _{j=1}^{k^{l-i}}\{b\ast k^{i}\}=b\ast k^{l}$ for  $1\le i\le l$. 

By scaling each  $\lambda _{i}\in P_{2}(b\ast k^{l-1})$ in the set  $\{(\lambda _{1},\lambda _{2},\ldots ,\lambda _{l-1})\}$ by a factor of $k$ to  $\{(k\ast \lambda _{1},k\ast \lambda _{2},\ldots ,k\ast \lambda _{l-1})\}$ , we get  $\sum _{j=1}^{k^{l-i}}\{b\ast k^{i}\}=b\ast k^{l}$ for  $1\le i\le l$ which are nothing but  $\{(\beta _{1},\beta _{2},\ldots ,\beta _{l-1})\}$ .

We use  $k$ instances of the girth maximum $(b\ast k^{l-1},l)$ BTU, and by choosing $p_{l+1}\in S_{b\ast k^{l}}$ so that partition between $p_{l}$  and  $p_{l+1}$  is $\sum _{j=1}^{1}\{b\ast k^{l}\}=b\ast k^{l}$ such that we get maximum girth by maximizing length of the micro-partition cycles, 

 $q_{l-1,1}=b\ast k^{l};y_{l-1}=k;q_{l,1}=b\ast k^{l+1};y_{l}=1$ and generalized Micro-partitions 
$x_{l,l-1,j,1}=k$ for $1\le j\le k$ and $x_{l-1,l,1,j}=k$ for $\le j\le k $.

Hence, the micro-partition cycles are $2\ast (k^{2}/k+1\text{--}1)=2\ast k$ and $2\ast (k/k+k\text{--}1)=2\ast k$ . We can show that the case where no micro-partition cycles arise leads to cycle length strictly less than $2\ast k$, and hence maximizing length of the micro-partition cycles leads to maximum girth. 

We hence obtain the $(b\ast k^{l},l+1)$ BTU with maximum girth which clearly lies in  $\Phi (\beta _{1},\beta _{2},\ldots ,\beta _{l})$. Hence the statement is true for  $r=l+1$ and hence by the principle of finite induction, the statement is true for all $r\ge 3;r\in \mathbb{N}$.

\subsection[Algorithm for girth maximizing BTU]{Algorithm for girth maximizing $(m,r)$ BTU}
{\textbf{Assumptions}: We assume that  $m\in \mathbb{N}$ is a composite number.}
\begin{enumerate}
\item We factorize  $m=k^{r-1}\ast b$  where  $k,b\in \mathbb{N}$  such that $b$  is minimized\textbf{.}
\item Choose $\alpha _{i}\in P_{2}(k^{i})$  as  $\sum _{j=1}^{1}\{k^{i}\}=k^{i}$ for  $1\le i\le r\text{--}1$ . 
\end{enumerate}
\begin{center}
\begin{tabular}{|m{1.0cm}|m{4.2cm}|}
\hline
~

\begin{equation*}
\alpha _{1}
\end{equation*}
 &
\begin{equation*}
\sum _{j=1}^{1}\{k\}=k
\end{equation*}
\\\hline
~

\begin{equation*}
\alpha _{2}
\end{equation*}
 &
\begin{equation*}
\sum _{j=1}^{1}\{k^{2}\}=k^{2}
\end{equation*}
\\\hline
\begin{equation*}
\ldots 
\end{equation*}
 &
\begin{equation*}
\ldots 
\end{equation*}
\\\hline
~

\begin{equation*}
\alpha _{r-1}
\end{equation*}
 &
\begin{equation*}
\sum _{j=1}^{1}\{k^{r-1}\}=k^{r-1}
\end{equation*}
\\\hline
\end{tabular}
\end{center}
The corresponding  $\gamma _{i}\in P_{2}(k^{r-1})$ are chosen as $\sum _{j=1}^{k^{r-1-i}}\{k^{i}\}=k^{r-1}$ for  $1\le i\le r\text{--}1$ . 
\begin{center}
\begin{tabular}{|m{1.0cm}|m{4.2cm}|}
\hline
~

\begin{equation*}
\gamma _{1}
\end{equation*}
 &
\begin{equation*}
\sum _{j=1}^{k^{r-2}}\{k\}=k^{r-1}
\end{equation*}
\\\hline
~

\begin{equation*}
\gamma _{2}
\end{equation*}
 &
\begin{equation*}
\sum _{j=1}^{k^{r-3}}\{k^{2}\}=k^{r-1}
\end{equation*}
\\\hline
\begin{equation*}
\ldots 
\end{equation*}
 &
\begin{equation*}
\ldots 
\end{equation*}
\\\hline
~

\begin{equation*}
\gamma _{r-1}
\end{equation*}
 &
\begin{equation*}
\sum _{j=1}^{1}\{k^{r-1}\}=k^{r-1}
\end{equation*}
\\\hline
\end{tabular}
\end{center}

The corresponding  $\beta _{i}\in P_{2}(m)$ are chosen as $\sum _{j=1}^{k^{r-1-i}}\{b\ast k^{i}\}=b\ast k^{r-1}=m$ for  $1\le i\le r\text{--}1$ . 
\begin{center}
\begin{tabular}{|m{1.0cm}|m{4.2cm}|}
\hline
~

\begin{equation*}
\beta _{1}
\end{equation*}
 &
\begin{equation*}
\sum _{j=1}^{k^{r-2}}\{b\ast k\}=b\ast k^{r-1}=m
\end{equation*}
\\\hline
~

\begin{equation*}
\beta _{2}
\end{equation*}
 &
\begin{equation*}
\sum _{j=1}^{k^{r-3}}\{b\ast k^{2}\}=b\ast k^{r-1}=m
\end{equation*}
\\\hline
\begin{equation*}
\ldots 
\end{equation*}
 &
\begin{equation*}
\ldots 
\end{equation*}
\\\hline
~

\begin{equation*}
\beta _{r-1}
\end{equation*}
 &
\begin{equation*}
\sum _{j=1}^{1}\{b\ast k^{r-1}\}=b\ast k^{r-1}=m
\end{equation*}
\\\hline
\end{tabular}
\end{center}


After computation of the optimal partitions, the search for a girth maximum BTU involves the following steps.
\begin{enumerate}
\item We construct a girth maximum  $(b\ast k,2)$ BTU.
\item We search for a girth maximum  $(b\ast k^{2},3)$  BTU.
\item We search for a  girth maximum  $(b\ast k^{3},4)$  BTU.
\item We search for a  girth maximum  $(b\ast k^{r-1},r)$ BTU.
\end{enumerate}
\subsection[\ Conjecture]{\textup{\ }Conjecture}
In order to construct a $(m,r)$ BTU with best girth where  $k^{r-1}=m/b;k=(m/b)^{1/{r-1}};k\in \mathbb{N}$ where  $m=k^{r-1}\ast b$ where  $k,b\in \mathbb{N}$ such that $b$ is minimized,  $\beta _{i}$ is chosen \ as $\sum _{j=1}^{k^{r-1-i}}\{b\ast k^{i}\}=b\ast k^{r-1}=m$ for  $1\le i\le r\text{--}1$, it is sufficient to construct $(k^{2},3)$ BTU with best girth, and use this as a template for making the rest of the connections as described by the following hierarchy of girth maximum BTUs

\begin{center}
\begin{tabular}{|m{1.0cm}|m{4.2cm}|m{1.8cm}|}
\hline
\begin{equation*}
i
\end{equation*}
 &
\begin{equation*}
\beta _{i}
\end{equation*}
 &
BTU\\\hline
~

\begin{equation*}
1
\end{equation*}
 &
\begin{equation*}
\sum _{j=1}^{k^{r-2}}b\ast k^{1}=m
\end{equation*}
 &
~

\begin{equation*}
(m,2)
\end{equation*}
\\\hline
~

\begin{equation*}
2
\end{equation*}
 &
\begin{equation*}
\sum _{j=1}^{k^{r-3}}\{b\ast k^{2}\}=b\ast k^{r-1}=m
\end{equation*}
 &
~

\begin{equation*}
(m,3)
\end{equation*}
\\\hline
~
 &
\begin{equation*}
\ldots 
\end{equation*}
 &
\begin{equation*}
\ldots 
\end{equation*}
\\\hline
~

\begin{equation*}
r-2
\end{equation*}
 &
\begin{equation*}
\sum _{j=1}^{k}\{b\ast k^{r-2}\}=b\ast k^{r-1}=m
\end{equation*}
 &
~

\begin{equation*}
(m,r-1)
\end{equation*}
\\\hline
~

\begin{equation*}
r-1
\end{equation*}
 &
\begin{equation*}
\sum _{j=1}^{1}\{b\ast k^{r-1}\}=b\ast k^{r-1}=m
\end{equation*}
 &
~

\begin{equation*}
(m,r)
\end{equation*}
\\\hline
\end{tabular}
\end{center}

\bigskip

\section[Search Problem for finding BTU with best girth where]{Search Problem for finding  $(b\ast k^{2},3)$ BTU with best girth where $k\in \mathbb{N}$ }
{\textup{For } $r=3$\textup{, with } $p_{1}=I_{b\ast k^{2}}$\textup{ and } $p_{2}\in S_{b\ast k^{2}}$\textup{ as per } $\Psi (\beta _{1})$ \textup{where } $\beta _{1}\in P_{2}(b\ast k^{2})$ \textup{refers to } $\sum _{j=1}^{k}\{b\ast k^{1}\}=b\ast k^{2}$\textup{, and} $\beta _{2}\in P_{2}(b\ast k^{2})$ \textup{refers to } $\sum _{j=1}^{1}\{b\ast k^{2}\}=b\ast k^{2}$\textup{ , to find } $p_{3}\in S_{b\ast k^{2}}$ \textup{such that the labeled BTU} $\{p_{1},p_{2},p_{3}\}$\textup{ has maximum girth among all } $(b\ast k^{2},3)$ \textup{BTUs.}}
\subsection[Algorithm to generate optimal partitions for a given value of  and]{Algorithm to generate optimal partitions for a given value of  $k$  and $r$ }
{The following algorithm generate optimal partitions  $\beta _{1},\beta _{2},\ldots ,\beta _{r\text{--}1}\in P_{2}(k^{r-1})$ for a given value of  $k$  and $r$ such that the girth maximum  $(k^{r-1},r)$  BTU lies in  $\Phi (\beta _{1},\beta _{2},\ldots ,\beta _{r\text{--}1})$ .}
 $m=k$ ;

for( $i=1;i\le r-1;i$ ++) \{

 $\beta _{i}$ refers to $\sum _{j=1}^{1}m$ ;

for( $z=1;z<i;z$ ++) \{

\  $k\ast \beta _{z}$ ; //scale partition $\beta _{z}$ by  $k$ 

\}

 $m=k\ast m$ ;

\}

\section[Search Problem for finding BTU with best girth where]{Search Problem for finding  $(b\ast k^{r-1},r)$ BTU with best girth where $k\in \mathbb{N}$}
 $p_{1}=I_{b\ast k^{2}}$ and  $p_{2}\in S_{b\ast k^{2}}$ as per  $\Psi (\beta _{1})$ where  $\beta _{1}\in P_{2}(b\ast k^{2})$ refers to  $\sum _{j=1}^{k^{r-2}}b\ast k^{1}=b\ast k^{r-1}$, $\beta _{2}\in P_{2}(b\ast k^{2})$ refers to  $\sum _{j=1}^{k^{r-3}}\{b\ast k^{2}\}=b\ast k^{r-1}$,  $\ldots $ and $\beta _{r-1}\in P_{2}(b\ast k^{2})$ refers to  $\sum _{j=1}^{1}\{b\ast k^{r-1}\}=b\ast k^{r-1}$, we need to find  $p_{3},\ldots ,\in p_{r-1}\in S_{b\ast k^{r-1}}$ such that the labeled BTU $\{p_{1},p_{2},p_{3},\ldots ,p_{r\text{--}1}\}$ has maximum girth among all  $(b\ast k^{r-1},r)$ BTUs. \ \ \ \ \ \ \ \ \ \ \ \ \ \ 

\section[Open Questions on girth maximum  BTU]{Open Questions on girth maximum  $(m,r)$  BTU}
\begin{enumerate}
\item What is the maximum attainable girth for a  $(m,r)$ BTU?
\item How do we construct an optimal search problem for finding a girth maximum $(m,r)$ BTU ?
\item What is the computational complexity of the search problem for finding a girth maximum $(m,r)$  BTU ?
\end{enumerate}
\section{CONCLUSION}
This paper describes the optimal partition parameters for a girth maximum  $(m,r)$ BTU. We mathematically prove results for optimal parameters $\beta _{1},\beta _{2},\ldots ,\beta _{r\text{--}1}\in P_{2}(m)$ such that the \ girth maximum $(m,r)$ BTU lies in $\Phi (\beta _{1},\beta _{2},\ldots ,\beta _{r\text{--}1})$ and create a framework for specifying a search problem for finding the girth maximum $(m,r)$ BTU. We also raise some open questions on girth maximum $(m,r)$  BTU. 

{\centering\scshape
References
\par}

\begin{enumerate}
\item Vivek S Nittoor and Reiji Suda, {\textquotedblleft}Balanced Tanner Units And Their Properties{\textquotedblright} , arXiv:1212.6882 [cs.DM].
\item Vivek S Nittoor and Reiji Suda, {\textquotedblleft}Parallelizing A Coarse Grain Graph Search Problem Based upon LDPC Codes on a Supercomputer{\textquotedblright}, Proceedings of 6\textsuperscript{th} International Symposium on Parallel Computing in Electrical Engineering (PARELEC 2011), Luton, UK, April 2011.
\item R. M. Tanner, {\textquotedblleft}A recursive approach to low complexity codes,{\textquotedblright} IEEE Trans on Information Theory, vol. IT-27, no.5, pp. 533-547, Sept 1981.
\item C.E. Shannon, {\textquotedbl}A Mathematical Theory of Communication{\textquotedbl},Bell System Technical Journal, vol. 27, pp.379-423, 623-656, July, October, 1948.
\item D. J. C. MacKay and R. M. Neal, {\textquotedblleft}Near Shannon limit performance of low density parity check codes,{\textquotedblright} Electron. Lett., vol. 32, pp. 1645--1646, Aug. 1996.
\item William E. Ryan and Shu Lin,{\textquotedblright}Channel Codes Classical and Modern{\textquotedblright}, Cambridge University Press, 2009.
\item F. Harary, Graph Theory, Addison-Wesley, 1969.
\item Frank Harary and Edgar M. Palmer, {\textquotedblleft}Graphical Enumeration{\textquotedblright}, Academic Press, 1973.
\item Martin Aigner, {\textquotedblleft}A course in Enumeration{\textquotedblright}, Springer-Verlag, 2007.
\end{enumerate}
\end{multicols}
\end{document}